\documentclass[]{revtex4}
\usepackage{hyperref}
\begin{document}

\title{Nilpotent Symmetries in Super-Group Field Cosmology}

 \author{ Sudhaker Upadhyay}
  \email{ sudhakerupadhyay@gmail.com;  sudhaker@iitk.ac.in}
\affiliation { Department of Physics, Indian Institute of Technology Kanpur, Kanpur 208016, India}

\begin{abstract}
 In this paper we study the  gauge invariance of the  third quantized super-group field cosmology 
   which is a model for multiverse.  Further, we propose both
  the infinitesimal (usual) as well as the finite superfield-dependent  BRST symmetry transformations  which leave the effective theory invariant.
 The  effects of finite superfield-dependent BRST transformations on the path integral  
 (so-called void functional in the case of third quantization)  are implemented.
 Within the finite superfield-dependent BRST formulation,
the finite superfield-dependent BRST transformations with specific parameter
switch  the void functional from one gauge  to another. We establish this
  result for the most general gauge with the help of explicit calculations
 which holds for all possible sets of gauge choices
 at both the classical and the quantum levels.   
 \end{abstract}

\maketitle 
\section{Introduction}
The biggest challenge for  all the  schemes that quantize gravity (in particular the 
background-independent schemes), 
beside obtaining a solid description of the fundamental
degrees of freedom for quantum spacetime, is to produce the testable
predictions \cite{aaa,do}. 
In the background-independent approaches to quantize gravity, the loop quantum gravity is 
a powerful candidate quantizing gravity in mathematically
rigorous and in non-perturbative way \cite{rov,thi}.
In this approach the Hamiltonian constraints are reformulated in terms of Ashtekar-Barbero  variables,  the densitized triad and the Ashtekar connection  \cite{as0, as, as1,as2,li,ge}.
These basic  variables,  after  the consideration  of a  holonomies of the
connection and the fluxes of the triads, are promoted to the  basic quantum operators.
  In the loop quantum gravity,  the 
complete quantum dynamics of spin network states are obtained by putting the quantum states into the larger framework of group field theories.
The group field theories are  basically the
field theories on group manifolds (or their Lie algebras) which  provide a
 background-independent
third quantized formalism for  gravity in any dimension and signature \cite{ori1,ori2}. 
In the group field theories, both the geometry and the topology are  dynamical  and theus described in purely
algebraic and combinatorial terms.
The Feynman diagrams of such theories have
an interpretation of the  spacetimes and  therefore the quantum amplitudes for these diagrams 
can be interpreted as   algebraic realization of a path
integral description of gravity  \cite{ori,per}.

  The topology of loop quantum gravity is fixed and hence its dynamics can be studied by 
  second quantization. However, the topology changing processes can not be analyzed by second quantization approach.
Therefore, to analyze the theories, which are described by topology changing processes, the ``third quantization" is mandatory \cite{bu,pi,pe,ma}. 
 Incidentally, the third quantization of loop quantum gravity leads to the group field theory \cite{ab, sm, ab1, ta}.
  The basic  idea behind the third quantization formalism is
 to treat the theory of multiverse as a quantum
field theory on superspace \cite{str}. However, the Wheeler-De Witt (minisuperspace) approximations of such group field theory are 
known as the group field cosmology  
\cite{aa, qi, gu, ew, qi1, cal, gl,sg}.

On the other hand, the supersymmetry has been proved as a prominent candidate for the dark matter \cite{sa}. 
Supersymmetry is also  important  in the study of  many phenomenological models beyond the standard model  
\cite{il,lm,wt,mg,mc}. Recently, a supersymmetric
 generalization for group field cosmology has  been made  which is known as super-group field cosmology 
 \cite{fai0}. The   super-group field cosmology is a gauge symmetric model of the multiverse 
 and hence the theory  contains  some spurious  degrees of freedom. To quantize the theory correctly one should remove these spurious
 degrees of freedom  by putting some well-defined constraint, called as the gauge-fixing, in the theory. 
To achieve this goal at quantum level a gauge-fixing term is added in the classical  action  of the theory.
 This gauge-fixing term reflects an introduction of Faddeev-Popov ghosts in the void  functional (the vacuum
 functional  of third quantization) which helps in defining the physical Hilbert state of the
 effective theory.  The supersymmetric 
 BRST transformation and the unitarity of super-group field cosmology has been studied recently  \cite{fai}.
Further,   the Slavnov-Taylor identity and renormalizability of the theory 
has been investigated \cite{sud}. 
 
 The generalization of usual BRST transformations, so-called the finite field-dependent BRST (FFBRST) transformation, has been studied in great details in the context of gauge theories \cite{sdj}. 
 The FFBRST transformations have found many important applications in the second quantized gauge field theories 
\cite{sb,sdj,sdj1,rb,rb1,smm,fs,sud1,susk, subp1,ssb,sudd,sud,fsm,rbs,rs}.
Therefore, it is worthwhile to study the FFBRST formulation in the case of
third quantized super-group field cosmology. 
With this motivation we feel that this is a glaring omission.

In the present paper we first review  the basic ideas of super-group field cosmology.
 Then, we discuss the importance of different gauge-fixing conditions and different sets of infinitesimal BRST transformation for such model of super-group field cosmology.
 The Jacobian of functional measure under such infinitesimal BRST transformations 
 has found constant.
 Further we generalize the third quantized BRST transformation by making
 the transformation parameter finite and  superfield-dependent
 because the theory resides on the superspace.
 After-that we define the supersource free void functional 
 for third quantized cosmological model. The functional measure of such void functional remains  
 unchanged under the third quantized infinitesimal (usual) BRST transformation, however, it
 does not remain invariant under finite superfield-dependent 
 BRST transformations. The  finite superfield-dependent BRST transformation
 amounts a change in the expression of void functional.
 To study such change we choose an specific  superfield-dependent parameter
 which helps to change the effective theory from one gauge to another.

The   paper is organized as follows. In section II, we discuss the third quantized supersymmetric
group cosmology. The different gauge conditions and BRST symmetries are studied in the section III. The
generalization of usual BRST symmetry of super-group field cosmology
is discussed in section IV. Furthermore, in   section V, we
relate the different gauge-fixing conditions of the theory under the effect of finite 
superfield-dependent BRST transformations.
The last section is reserved for results and conclusions. 
\section{Supersymmetrization of group field cosmology}
In this section we provide a description within loop quantum cosmology of the spatially ﬂat, homogeneous and isotropic universe   with a massless scalar field  as matter.
 The four-dimensional  metric described in terms of three metric $q_{ab}$ is then defined by
\begin{eqnarray}
ds^2=-N^2 (t)dt^2+a^2(t)q_{ab}dx^adx^b,
\end{eqnarray}
where $N(t)$ and $a(t)$ are lapse function and scale factor respectively. Here Latin indices $a$ and$b$ denote the spatial indices.
 In loop quantum gravity  the phase space is described by a $SU(2)$ gauge connection, the Ashtekar-Barbero connection
 $A_a^A$
 and its canonically conjugate momentum and the densitized triad  $E_A^a$ which plays the
role of an ``electric field". The capital alphabets $A,B,...,$  are $SU(2)$ indices and label 
new degrees of freedom introduced when passing to the triad formulation. To define these 
objects, first one introduces the co-triad
$e_a^A$ defined
as $q_{ab} = e_a^Ae_b^B\delta_{AB}$, where $\delta_{AB}$ stands for the Kronecker delta in 
the three dimensions.
Then we define the triad, $e_A^a$, as its inverse $e_A^ae^B_b=\delta^a_b\delta_A^B$.
Now, the Ashtekar-Barbero connection reads $A_a^A=\Gamma_a^A+\gamma K_a^A$,
where $\gamma$
is the Barbero-Immirzi parameter and $K_a^A$ is the extrinsic curvature in triadic
form, and $\Gamma_a^A$ is the spin connection compatible with the densitized triad.
The curvature of connection $A^i_a$ in the loop quantum cosmology is expressed through the holonomy around a loop such that
the area of a loop   cannot be smaller than a fixed minimum area because
the smallest eigenvalue of the area operator in loop quantum gravity is nonzero.
Now, one defines the  eigenstates of the volume operator (finite cell) $\cal{V}$  with a basis, $|\nu \rangle $,  as follows:  
$ {\cal{V}} |\nu \rangle = 2 \pi \gamma G |\nu| |\nu \rangle
$, where gravitational conﬁguration variable $\nu = \pm a^2 {\cal{V}}_0 /2\pi \gamma G$ has the dimensions of length. 
 The Hamiltonian  constraint in the Plank units  for a homogeneous isotropic universe
 is defined as \cite{fai0}
\begin{equation} 
 - B(\nu)[E^2 - \partial^2_\phi] \Phi(\nu, \phi)= K ^2\Phi(\nu, \phi)  =0,
\end{equation}
  where $\Phi(\nu, \phi)$ is a wavefunction on configuration space
  and $E^2 $ is a difference operator defined as
   \begin{eqnarray}
 -E^2 \Phi(\nu, \phi) =  \frac{C^+(\nu)}{ B(\nu)} \Phi(\nu+\nu_0 , \phi)
+\frac{C^0(\nu)}{ B(\nu)}\Phi(\nu, \phi)  +\frac{C^-(\nu)}{ B(\nu)}\Phi(\nu-\nu_0, \phi),
\end{eqnarray}
and $K ^2= - B(\nu)[E^2 - \partial^2_\phi] $.
Here $\nu_0$ is an elementary length
unit, usually defined by the square root of the area gap   and the functions $B(\nu), C^+(\nu),
C^0(\nu)$ and $C^-(\nu)$ that
  depend
on the choice of the lapse function and on  the details of quantization scheme.
For example,  in an improved dynamic scheme, these 
functions for particular choice of lapse function, i.e. $N=1$,  have the following form \cite{aste}:
\begin{eqnarray}
B(\nu)&=& \frac{3\sqrt{2}}{8\sqrt{\sqrt{3}\pi\gamma G}}|\nu | \left| \left|\nu +
\frac{\nu_0}{4}\right|^{\frac{1}{3}} -\left|\nu -
\frac{\nu_0}{4}\right|^{\frac{1}{3}}\right|^3,\nonumber\\
C^+(\nu)&=& \frac{1}{12\gamma\sqrt{2\sqrt{3}}}\left|\nu +\frac{\nu_0}{2}\right| \left| \left|\nu +
\frac{\nu_0}{4}\right|  -\left|\nu +
\frac{3\nu_0}{4}\right|\right|,\nonumber\\
C^0(\nu) &=& -C^+(\nu) -C^+(\nu -\nu_0),\nonumber\\
C^-(\nu)&=&  C^+(\nu-\nu_0).\label{fun1}
\end{eqnarray}

Now, the classical actions for the bosonic   group field cosmology 
 is defined by \cite{cal, fai0}
 \begin{equation}
 S_{bose} =\sum_\nu \int d\phi\ {\cal L}_{bose}= \sum_\nu \int d\phi \, \,    \Phi (\nu, \phi)  K^2  \Phi(\nu, \phi). 
\end{equation}
It is worthwhile to  analyse the fermionic distribution of universes also which might
lead  
to the correct  value of the cosmological constant. Since  the significant value
of cosmological constant is not obtained by considering  only bosonic distributions of 
the universes in the multi-universe scenario \cite{kolm}. Keeping this point in mind, 
the free action corresponding to the fermionic  group field cosmology is constructed 
as \cite{fai0}
\begin{equation}
 S_{fermi} = \sum_\nu \int d\phi \, \,  \Psi^j(\nu, \phi) K_j^i \Psi_i(\nu, \phi),
\end{equation}
 where $\Psi_i (\nu, \phi) = (\Psi_1(\nu, \phi), \Psi_2(\nu, \phi) )$ is a fermionic spinor 
 field and   $K_{ij} = (\gamma^\mu )_{ij}K_\mu$ is an operator. The spinor indices are raised
and lowered by the second-rank antisymmetric tensors $C^{ij}$ and $C_{ij}$ respectively.
These tensors satisfy following condition $C_{ij}C^{il} = \delta^l_j$.
The above bosonic and the fermionic actions describe the bosonic and the fermionic universes
in the multiverse and hence, it is worthwhile to construct a supersymmetric gauge invariant model describing
multiverse. For this purpose we define two complex scalar super-group fields
$\Omega(\nu, \phi, \theta) $ and $ \Omega^{\dagger}  (\nu, \phi, \theta)$ and a 
spinor super-group field $\Gamma_a (\nu, \phi, \theta)$, which are suitably contracted with generators 
 of a Lie algebra, $[T_A, T_B] = i f_{AB}^C T_C$, as
\begin{eqnarray}
   \Omega(\nu, \phi, \theta) &=&\Omega^A(\nu, \phi, \theta) T_A, \nonumber \\ 
\Omega^{\dagger}(\nu, \phi, \theta) &=& \Omega^{\dagger A}(\nu, \phi, \theta)  T_A \nonumber \\ 
\Gamma_a (\nu, \phi, \theta) &=& \Gamma_a^A (\nu, \phi, \theta)T_A. 
\end{eqnarray}
The extra variable $\theta$ is Grassmannian in nature which defines the extra direction in superspace.
The super-covariant derivatives of these superfields are defined by
\cite{fai0}
\begin{eqnarray}
  \nabla_a  \Omega^A(\varrho)&=& D_a\Omega^A(\varrho) -i f_{CB}^A\Gamma^C_a(\varrho) \Omega^B(\varrho),\nonumber\\
\nabla_a \Omega^{A \dagger}(\varrho)  &=& D_a \Omega^{A \dagger}(\varrho)  
+ i  f_{CB}^A\Omega^{C \dagger}(\varrho)  \Gamma^B_a (\varrho), 
\end{eqnarray}
where super-derivative $ D_a = \partial_a + K^b_a \theta_b$ and superspace variables $(\nu, \phi, \theta) :=\varrho$.
We define the field-strength for  a matrix valued spinor field ($ \Gamma^A_a $) as follows
$
 \omega^A_a (\varrho) = \nabla^b \nabla_a \Gamma^A_b(\varrho) $.
 
Now, we are  able to construct 
 the classical action for the super-group field cosmology as \cite{fai0}
 \begin{eqnarray}
S_{0} =\sum_\nu \int d\phi \, \,    \left[D^2 \{\Omega_A^{\dagger} (\varrho) \nabla_a^2 \Omega^A
 (\varrho) 
  + \omega_A^a (\varrho) \omega^A_a(\varrho)  \}\right]_|,
\end{eqnarray}
where $'|'$ refers $\theta_a = 0$ at the end of calculation.  
This classical action remains invariant  under following gauge transformations
\begin{eqnarray}
  \delta \Omega^A(\varrho) &=&  if_{CB}^A\Lambda^C (\varrho)\Omega^B(\varrho) ,\nonumber\\
\delta \Omega^{A \dagger}(\varrho)  &=& -i f^A_{CB}\Omega^{C\dagger}(\varrho) \Lambda^B(\varrho), \nonumber\\
 \delta \Gamma^A_a(\varrho) &=& \nabla_a \Lambda^A(\varrho).
\end{eqnarray}
where the bosonic transformation parameter $\Lambda^A$ is   infinitesimal in nature. 
\section{The infinitesimal BRST symmetries for different gauge conditions} 
In this section, we will construct the infinitesimal  nilpotent BRST symmetries for the theory. For this purpose
we  need to fix a gauge before quantizing the theory as the theory is gauge invariant and therefore possesses some redundant degrees of freedom. The
general gauge-fixing condition for the theory is given by
\begin{equation}
 F^A[ \Gamma^A_a (\varrho) ] =0.
\end{equation}
This can be incorporated at a quantum level by adding an
appropriate gauge-fixing term to the classical action. 
The linearized gauge-fixing term  with the help of
Nakanish-Lautrup auxiliary superfield  $  B^A( \nu, \phi, \theta) $ is given by
\begin{equation}
 S_{gf} = \sum_\nu \int d\phi \, \,   \left[D^2\{ B_A (\varrho)F^A[ \Gamma^A_a (\varrho) ]\}\right]_|.\label{gf}
\end{equation}
Since the gauge-fixing condition produces the Faddeev-Popov ghost term in the effective theory. Therefore, in this case, 
the ghost term induced   by  (\ref{gf}) is 
\begin{equation}
  S_{gh} = \sum_\nu \int d\phi \, \,   
 \left[D^2\{\bar c_A(\varrho)s_b F^A[ \Gamma^A_a (\varrho) ]\}\right]_|,\label{gh}
\end{equation}
where $   {c}^A(\varrho)  $
and   $ \bar{c}^A (\varrho) $ are   
the ghost and anti-ghost superfields respectively, however, 
$s_b$ denotes the Slavnov variation.
Now, the total effective action for super-group field cosmology having general gauge choice is written by
\begin{equation}
S_T =  S_{0}+  S_{gh}+  S_{gf}.\label{full}
\end{equation}
However, for
specific choice (say covariant gauge choice) $F^A=D^a\Gamma^A_a(\varrho) =0$,
the above action $S_T$ 
reduces to  \cite{fai}:
\begin{eqnarray}
S_T =  S_{0}+  \sum_\nu \int d\phi \, \,   \left[D^2\{ B_A (\varrho)D^a\Gamma^A_a(\varrho)+ \bar c_A(\varrho) D^a\nabla_a c^A (\varrho) ] \}\right]_|,\label{ori}
\end{eqnarray} 
which
remains invariant under 
following   third quantized infinitesimal BRST transformations \cite{fai} 
\begin{eqnarray}
  \delta_b\,\Omega^A(\varrho)&= &if_{CB}   ^Ac^C (\varrho)\Omega^B(\varrho) \ \delta\lambda, \nonumber \\
\delta_b\,  \Omega^{A \dagger}(\varrho)  &= & -i f^A_{CB}\Omega^{ \dagger C}(\varrho)c^B(\varrho)\ \delta\lambda, 
\nonumber \\ 
\delta_b\, c^A( \varrho)&= &f^A_{CB}c^{  C}(\varrho) c^B(\varrho)\ \delta\lambda, 
\nonumber \\
\delta_b\, \Gamma^A_a (\varrho) &= & \nabla_a c^A(\varrho)\ \delta\lambda, 
\nonumber \\ 
\delta_b\, \bar{ c} ^A(\varrho) &= & B^A(\varrho)\ \delta\lambda,
 \nonumber \\
\delta_b\, B^A(\varrho) &= &0,\label{brs}
\end{eqnarray}
where $  \delta\lambda$ is an infinitesimal, anticommuting and global parameter. 
These transformations are nilpotent in nature,  i.e.,  $\delta_b^2 =0$. 
Utilizing the above BRST transformation we are able to write the sum of gauge-fixing and ghost parts of the action given in (\ref{gf}) and (\ref{gh})
in terms of BRST variation of gauge-fixed fermion  as follows
\begin{eqnarray}
S_{gf}+S_{gh} = \sum_\nu\int d\phi \ s_b\left [D^2 \{\bar{c}_A  (\varrho) F^A[ \Gamma^A_a (\varrho) ] \}\right]_|.
\end{eqnarray}
Furthermore, to analyse the theory in  massless Curci-Ferrari gauge (which is a non-linear 
gauge) we make the following shift in auxiliary superfield:
$B^A( \nu, \phi, \theta) \longrightarrow B^A(\varrho)  -\frac{1}{2}f^A_{BC} \bar c^B (\varrho) 
c^C(\varrho)$. Performing such shift of auxiliary superfield
the total effective action  in non-linear gauge is 
described by
\begin{eqnarray}
S_T &= &S_0+ \sum_\nu \int d\phi \, \,   
 \left[D^2\left\lbrace B_A (\varrho) D^a \Gamma^A_a (\varrho)+\frac{1}{2}\bar c_A(\varrho) D^a \nabla_a c^A
(\varrho) \right.\right.\nonumber\\
&+&\left.\left.\frac{1}{8} f_{BC}^Af_{A}^{GH}\bar c^B (\varrho)c^C(\varrho) \bar c_G(\varrho) c_H(\varrho)\right\rbrace\right]_|.\label{acta}
\end{eqnarray}
This   effective action
(\ref{acta}) admits the  fermionic rigid 
 non-linear BRST invariance. The BRST transformation  characterized by
    infinitesimal parameter $\delta\lambda$ is given by
\begin{eqnarray} 
  \delta_b\,\Omega^A(\varrho)&= &if_{CB}
  ^Ac^C (\varrho)\Omega^B(\varrho) \ \delta\lambda, \nonumber \\
\delta_b\,  \Omega^{A \dagger}(\varrho)  &= & -i f^A_{CB}\Omega^{ \dagger C}(\varrho)c^B(\varrho)\ \delta\lambda, 
\nonumber \\ 
\delta_b\, c^A(\varrho)&= &f^A_{CB}c^{  C}(\varrho) c^B(\varrho)\ \delta\lambda, 
\nonumber \\
 \delta_b\, \Gamma^A_a (\varrho) &= & \nabla_a c^A(\varrho)\ \delta\lambda, 
\nonumber \\ 
\delta_b\, \bar{ c} ^A(\varrho) &= & B^A(\varrho)\ \delta\lambda -\frac{1}{2}f^A_{BC} \bar c^B (\varrho) 
c^C(\varrho)\ \delta\lambda,
 \nonumber \\
\delta_b\, B^A(\varrho) &= & -\frac{1}{2}f^A_{BC}  c^B (\varrho) 
B^C(\varrho)\ \delta\lambda \nonumber\\
&-&\frac{1}{8} f_{BC}^Af_{GH}^{C}c^B(\varrho) c^G(\varrho) \bar c^H
(\varrho)\ \delta\lambda,\label{brs1}
\end{eqnarray}
which is also nilpotent in nature, i.e., $\delta_b^2 =0$.

To study the quantum effects for third quantize super-group field cosmology
 we first  define the source free void functional  
as follows:
\begin{eqnarray}
\left\langle 0|0\right\rangle =Z[0]=\int {\cal D} M e^{iS_T(\varrho)},\label{gen}
\end{eqnarray}
where  ${\cal D} M\equiv {\cal D}\Omega{\cal D}\Omega^\dag{\cal D} \Gamma_a
{\cal D} B{\cal D}c{\cal D}\bar c$ is the path integral measure.
This path integral measure remains invariant under the infinitesimal BRST transformation
given in(\ref{brs}) because the Jacobian of path integral measure, ${\cal D} M$,
 for  such BRST transformations  comes unit. 
\section{Finite superfield-dependent BRST symmetry for super-group field cosmology }
 In this section, we construct the  finite  superfield-dependent BRST transformations for the  
 third quantized super-group field cosmology.
 The properties of the usual BRST transformation  do not depend on whether the parameter $\delta\lambda$
  is (i) finite or infinitesimal, (ii) superfield-dependent or not, as long as it is anticommuting
and global. These observations give us a freedom to generalize the BRST transformation by making the parameter,
 $\delta\lambda$  finite and superfield-dependent without affecting its other properties.

  In order to  do that 
we first  make all the generic superfields $\Phi_i(\varrho, \kappa) = \Phi_i^{A}(\varrho, \kappa) T_A $, where $\Phi_i^{  A}
 = (\Omega^A, \Omega^{\dag A},  \Gamma^A_a, B^A, c^A, \bar c^A)$, 
   to depend on  a continuous parameter, $\kappa: 0\le \kappa \le 1$, in 
such a manner that ${\Phi_i}   (\varrho, 0 )$ are the initial superfields and
 $ {\Phi_i}   (\varrho, 1)$ are the transformed superfields. 
We also define a functional $\Theta [{\Phi_i (\varrho)}   ]$  with odd Grassmann parity. 
Now, we make the  infinitesimal parameter in the  BRST transformation superfield dependent and hence the 
infinitesimal superfield-dependent BRST 
transformation takes the following form \cite{sdj}:
\begin{equation}
\frac{ d}{d \kappa}{\Phi_i}   (\varrho, \kappa ) =  s_b  {\Phi_i}    (\varrho, \kappa )\
\epsilon [{\Phi}_i   (\varrho,\kappa )],
\label{dif}
\end{equation}
where $\epsilon [{\Phi}_i  (\varrho,\kappa )]$ is an infinitesimal superfield-dependent parameter
and $s_b {\Phi_i} $ is the BRST variation of superfields without parameter known as the Slavnov variation. By 
integrating these equations 
from $ \kappa=0$ to $\kappa=1$, it is shown  
that the ${\Phi_i} (\varrho, 1) $ are related to ${\Phi_i}   (\varrho, 0)
$ by the  finite superfield-dependent BRST transformation  as follows
\begin{equation}
{\Phi^i} (\varrho, 1) = {\Phi_i}   (\varrho, 0) + s_b  {\Phi_i}    (\varrho, 0) \Theta [{\Phi_i}   
(\varrho)],
\end{equation}
where 
\begin{eqnarray}
\Theta[{\Phi_i}   (\varrho)]= \int_0^1 d\kappa' \epsilon [{\Phi}_i (\varrho,\kappa' )].
\label{fin}
\end{eqnarray}
Consequently, the  finite superfield-dependent version of
 BRST transformation (\ref{brs}) 
 for the super-group field cosmology in linear gauge  is demonstrated by
\begin{eqnarray}
f\,\Omega^A(\varrho)&= &if_{CB}   ^Ac^C (\varrho)\Omega^B(\varrho) \ \Theta[\Phi_i], \nonumber \\
f\,  \Omega^{A \dagger}(\varrho)  &= & -i f^A_{CB}\Omega^{ \dagger C}(\varrho)c^B(\varrho)\  \Theta[\Phi_i], 
\nonumber \\ 
f\, c^A(\varrho)&= &f^A_{CB}c^{  C}(\varrho) c^B(\varrho)\  \Theta[\Phi_i], 
\nonumber \\
f\, \Gamma^A_a (\varrho) &= & \nabla_a c^A(\varrho)\  \Theta[\Phi_i], 
\nonumber \\ 
f\, \bar{ c} ^A(\varrho) &= & B^A(\varrho)\  \Theta[\Phi_i],
 \nonumber \\
f\, B^A(\varrho) &= &0,\label{ffbrs}
\end{eqnarray}
The above finite BRST symmetry transformation is a symmetry of the effective action (\ref{ori}) only but not of
 the corresponding functional measure defined in the Eq. (\ref{gen})
because the Jacobian for path integral measure  in the expression of void functional  does not appear as a constant factor due to 
superfield-dependent nature of transformation parameter. 
Under such transformation the
Jacobian changes as  
$
{\cal D}M    =J[{\Phi}_i   (\varrho,\kappa)] {\cal D}M   (\varrho,\kappa)   
$, where Jacobian depends on superfields.        
It has been shown  that this nontrivial Jacobian can be replaced within the 
functional integral as
\begin{equation}
J[{\Phi_i}   (\varrho, \kappa)] \rightarrow e^{iS_1[{\Phi}_i   (\varrho, \kappa)]},
\end{equation}
where $S_1[{\Phi}_i   (\varrho, \kappa)]$ is some local functional (here local stands for superfield dependent). The  Jacobian $J[{\Phi_i}]$
can be incorporated in the functional integral without 
changing the physical theory if and only if  
\cite{sdj}
\begin{eqnarray}
 \frac{1}{J [{\Phi}_i   (\varrho, \kappa)]}\frac{d J [{\Phi}_i   (\varrho, \kappa)]}{d\kappa} -i\frac{dS_1[{\Phi}_i   (\varrho, \kappa)]}{d\kappa}  =0.
\label{mcond}
\end{eqnarray}
Following the formulation given in Ref. \cite{sdj}, we calculate the 
infinitesimal change in Jacobian with the help of
 the following expression, 
 \begin{eqnarray} 
 \frac{1}{J [{\Phi}_i   (\varrho, \kappa)]}\frac{dJ [{\Phi}_i   (\varrho, \kappa)]}{d\kappa}&=& 
 -\sum_\phi \int d\phi  \,\,      \left[ -s_b  \Omega^A (\varrho, \kappa)
\frac{ \delta \epsilon [\Phi_i (\varrho, \kappa)]}{\delta
\Omega^A(\varrho, \kappa)} \right. \nonumber \\ &&\left.  -s_b \Omega^{\dag A} (\varrho, \kappa)
\frac{ \delta \epsilon [\Phi_i (\varrho, \kappa)]}{\delta
\Omega^{\dag A}(\varrho, \kappa)} +  s_b  \Gamma^A_a (x) \frac{ \delta \epsilon [\Phi_i (\varrho, \kappa)]}{\delta
\Gamma^A_a (\varrho, \kappa)} \right. \nonumber \\ &&\left.  -s_b  c^A (\varrho, \kappa)
\frac{ \delta \epsilon [\Phi_i (\varrho, \kappa)]}{\delta
c^A(\varrho, \kappa)}  - s_b  \bar c^A (\varrho, \kappa)
\frac{ \delta \epsilon [\Phi_i (\varrho, \kappa)]}{\delta
\bar c^A (\varrho, \kappa)}  \right. \nonumber \\ &&\left.  +s_b B^A(\varrho, \kappa)
\frac{ \delta \epsilon [\Phi_i (\varrho, \kappa)]}{\delta
B^A(\varrho, \kappa)} \right]_|.\label{jaceva}
\end{eqnarray}
Therefore, the Jacobian under finite superfield-dependent BRST
 transformation extends the effective action $S_T(\varrho)$ within the
  void functional  by  a terms $S_1(\varrho))$   as follows:
 \begin{eqnarray}
 Z[0]\left(\equiv\int {\cal D} M e^{iS_T(\varrho)} \right)\longrightarrow 
 Z'[0]\left(\equiv\int {\cal D} M e^{i(S_T(\varrho) +S_1
 (\varrho))} \right),
 \end{eqnarray}
 where $S_T+S_1$ is an extended effective action. In the next section,
 we will elaborate this in more detailed way.
 \section{Effects of finite symmetry on the third quantized path integral}
 In this section, we  discuss the effect of finite superfield-dependent BRST
 transformation on functional measure given in Eq. (\ref{gen})
 for a particular choice of finite superfield-dependent parameter.
 In this regard we establish a connection between
 two different but arbitrary gauge choices of super-group field cosmology.
 For this purpose,   we first identify an appropriate superfield-dependent parameter $\Theta[\Phi_i]$ 
 involved in the Eq. (\ref{ffbrs}). In this case  $\Theta[\Phi_i]$ is obtainable 
 from  the following infinitesimal superfield-dependent parameter
 \begin{eqnarray}
\epsilon [\Phi_i (\varrho, \kappa)] =-i\sum_\phi \int d\phi\ D^2\left[\bar c_A(\varrho,\kappa )  
  F_1^A[\Gamma^A(\varrho, \kappa) ]
 -\bar c_A(\varrho,\kappa )  F_2^A[\Gamma^A(\varrho, \kappa)]   \right]_|,
 \end{eqnarray}
 using the relation (\ref{fin}). Here $F_1^A[\Gamma^A(\varrho, \kappa) ] $ and
  $F_2^A[\Gamma^A(\varrho, \kappa) ] $ are two 
 arbitrary gauge-fixing conditions for the theory
 of super-group field cosmology.
 
Using the expression (\ref{jaceva}),  the change in Jacobian for the above $\epsilon [\Phi_i (\varrho, \kappa)]$ is calculated by
 \begin{eqnarray}
\frac{1}{J}\frac{dJ}{d\kappa} &=&i\sum_\phi \int d\phi\    D^2\left[ -
 B_{A}(\varrho, \kappa) \lbrace F_1^A[\Gamma^A(\varrho, \kappa) ] 
 -F_2^A[\Gamma^A(\varrho, \kappa) ]\rbrace
\right.
\nonumber\\
 &+&\left.\lbrace s_b F_1^A[\Gamma^A(\varrho, \kappa) ] 
 -s_bF_2^A[\Gamma^A(\varrho, \kappa) ]\rbrace \bar c_{A} (\varrho, \kappa)\right]_|,\nonumber\\
 &=&\sum_\phi \int d\phi\  D^2\left[- 
 B_{A}(\varrho, \kappa) \lbrace F_1^A[\Gamma^A(\varrho, \kappa) ] 
 -F_2^A[\Gamma^A(\varrho, \kappa) ]\rbrace
\right.
\nonumber\\
 &-&\left.  \bar c_{A} (\varrho, \kappa)\lbrace s_b F_1^A[\Gamma^A(\varrho, \kappa) ] 
 -s_bF_2^A[\Gamma^A(\varrho, \kappa) ]\rbrace\right]_| \label{jc1}
\end{eqnarray}
The Jacobian $J$ can be written as $ e^{iS_1[{\Phi}_i   (\varrho, \kappa)]}$ 
  when the
condition (\ref{mcond}) is satisfied. We make the following
ansatz for the functional $S_1$ in this case:
\begin{eqnarray}
S_1[{\Phi}_i   (\varrho, \kappa)]  &=&\sum_\phi \int d\phi\  \left[D^2 \lbrace \xi_1(\kappa)
 B_{A}(\varrho, \kappa)  F_1^A[\Gamma^A(\varrho, \kappa) ] 
 \right.\nonumber\\
 &+& \left.\xi_2(\kappa)
 B_{A}(\varrho, \kappa)  F_2^A[\Gamma^A(\varrho, \kappa) ] 
\right.\nonumber\\
 &+&\left. \xi_3(\kappa ) \bar c_{A} (\varrho, \kappa) s_b F_1^A[\Gamma^A(\varrho, \kappa) ] 
\right.\nonumber\\
 &+&\left.  \xi_4(\kappa ) \bar c_{A} (\varrho, \kappa) s_bF_2^A[\Gamma^A(\varrho, \kappa) ] \rbrace\right]_|,\label{s1}
\end{eqnarray}
where $\xi_1(\kappa), \xi_2(\kappa), \xi_3(\kappa)$ and $\xi_4(\kappa)$
are arbitrary  $\kappa$-dependent constants which satisfy
following boundary conditions
\begin{eqnarray}
\xi_1(\kappa =0)= \xi_2(\kappa =0)=\xi_3(\kappa =0)=\xi_4(\kappa =0)=0.\label{bd}
\end{eqnarray} 

Further,  equations (\ref{jc1}), (\ref{s1}) and (\ref{mcond}) yields
\begin{eqnarray}
&&\sum_\phi \int d\phi\ \left[D^2 \lbrace  (\xi_1' +1)
 B_{A}(\varrho, \kappa)  F_1^A[\Gamma^A(\varrho, \kappa) ]
 \right.\nonumber\\
 &+& \left. (\xi_2' -1) 
 B_{A}(\varrho, \kappa)  F_2^A[\Gamma^A(\varrho, \kappa) ] 
\right.\nonumber\\
 &+&  \left. (\xi_3' +1) \bar c_{A} (\varrho, \kappa) s_b F_1^A[\Gamma^A(\varrho, \kappa) ] 
\right.\nonumber\\
 &+& \left. (\xi_4' -1)  \bar c_{A} (\varrho, \kappa) s_bF_2^A[\Gamma^A(\varrho, \kappa) ]
 \right.\nonumber\\
 &+&  \left.(\xi_1 -\xi_3 ) B_{A} (\varrho, \kappa) s_b F_1^A[\Gamma^A(\varrho, \kappa)]
\epsilon [\Phi_i (\varrho, \kappa)] \right. \nonumber\\
 &+& \left. (\xi_2 -\xi_4 ) B_{A} (\varrho, \kappa) s_b F_2^A[\Gamma^A(\varrho, \kappa)]
\epsilon [\Phi_i (\varrho, \kappa)] 
 \rbrace\right]_|
 =0,
\end{eqnarray}
where prime denotes the differentiation w.r.t. $\kappa$.
Now, equating the coefficients of L.H.S. and R.H.S. of the above equation, we get the 
following differential 
equations 
\begin{eqnarray}
\xi_1' +1=0,\ \ \xi_2' -1=0,\ \  \xi_3' +1=0,\ \  \xi_4' -1=0,\label{de}
\end{eqnarray}
which satisfy the condition, $\xi_1 -\xi_3=\xi_2 -\xi_4=0$.
The solutions of the differential equations given in (\ref{de}), satisfying  boundary conditions
mentioned in Eq. (\ref{bd}), are
\begin{eqnarray}
\xi_1(\kappa) =-\kappa,\ \ \xi_2(\kappa) = \kappa,\ \ \xi_3(\kappa) =-\kappa,\ \ \xi_4(\kappa) = \kappa.  
\end{eqnarray}
Now, plugging these values back in Eq. (\ref{s1}), the expression of
$S_1$ precisely becomes
\begin{eqnarray}
S_1[{\Phi}_i   (\varrho, \kappa)]  &=&\sum_\phi \int d\phi\ \left[D^2 \lbrace -\kappa
 B_{A}(\varrho, \kappa)  F_1^A[\Gamma^A(\varrho, \kappa) ] 
 \right.\nonumber\\
 &+&\left.\kappa
 B_{A}(\varrho, \kappa)  F_2^A[\Gamma^A(\varrho, \kappa) ] 
\right.\nonumber\\
 &-& \left.\kappa \bar c_{A} (\varrho, \kappa) s_b F_1^A[\Gamma^A(\varrho, \kappa) ] 
\right.\nonumber\\
 &+& \left.\kappa\bar c_{A} (\varrho, \kappa) s_bF_2^A[\Gamma^A(\varrho, \kappa) ] \rbrace\right]_|,
\end{eqnarray}
which vanishes at $\kappa =0$. However,  at $\kappa =1$ (under finite superfield-dependent BRST transformation)   the void
functional $Z[0]$ defined with 
gauge condition $F_1^A[ \Gamma^A_a (\varrho) ]=0$ transforms as follows:
\begin{eqnarray}
Z[0]\left(\equiv\int {\cal D} M e^{iS_T}\right) &\stackrel{ finite\ BRST  }{-----\longrightarrow}&  Z'[0]
\left(\equiv\int J( =e^{iS_1|_{\kappa=1}}){\cal D} M e^{iS_T}\right)\nonumber\\
& &= \int  {\cal D} M e^{i(S_T+S_1|_{\kappa=1})},
\end{eqnarray}
where $S_T$ refers to the total effective action for the gauge condition $F_1^A[ \Gamma^A_a (\varrho) ]=0$ as
\begin{eqnarray}
S_T =S_0 + \sum_\nu \int d\phi \, \,   \left[D^2\{ B_A (\varrho)F_1^A[ \Gamma^A_a (\varrho) ]
+\bar c_A(\varrho)sF_1^A[ \Gamma^A_a (\varrho) ]\}\right]_|, 
\end{eqnarray} and hence $S_T+S_1|_{\kappa=1}$ becomes
\begin{eqnarray}
S_T+S_1|_{\kappa=1}=S_0 + \sum_\nu \int d\phi \, \,   \left[D^2\{ B_A (\varrho)F_2^A[ \Gamma^A_a (\varrho) ]
+\bar c_A(\varrho)s_bF_2^A[ \Gamma^A_a (\varrho) ]\}\right]_|,
\end{eqnarray}
which is nothing but the complete effective actions corresponding to another set of
  gauge choice   $F_2^A[ \Gamma^A_a (\varrho) ]=0$.
Consequently, $Z'[0]$ refers to the 
 void functional of third quantized super-group field cosmology defined for  different 
gauge-fixing
 condition
  $F_2^A[ \Gamma^A_a (\varrho) ]=0$.
More concretely, the finite superfield-dependent BRST transformation 
with  parameter $\epsilon [\Phi_i (\varrho, \kappa)] =i \sum_\phi \int d\phi \left[ f^A_{BC} \bar c_A \bar c^B c^C \right]_|$  
will connect  the effective actions  of super-group field cosmology in linear and non-linear gauges.
  Hence, remarkably, we conclude that
  the finite super-field dependent transformations with specific 
  transformation parameter map two different gauges of super-group field cosmology  
  which will certainly help in explaining the different pole structures
  of propagators of the theory.
\section{Conclusion}
In this paper we have discussed  the   supersymmetric group field cosmology which
is a model for homogeneous and isotropic  multiverse.
 In the multiverse scenario, the gauge and the matter fields describe the
different universes.  
Furthermore, we have constructed the third quantized infinitesimal BRST transformations
of the super-group field cosmology defined for the most
general gauge-fixing condition.  
The finite superfield-dependent BRST symmetries, characterized by
an arbitrary superfield-dependent parameter, have  been demonstrated by integrating the infinitesimal BRST transformation  of the
 super-group field cosmology. Within the formalism, we have defined the
void functional (so-called the generating functional in the case of second quantized field theories) for this third 
quantized  super-group field cosmology. 
The effects of infinitesimal and finite superfield-dependent of BRST
transformations on the functional measure of
 void functional have also been reported. Within the analyses,  we have found that
the infinitesimal BRST transformation leaves  both the effective action and  the
 functional measure
invariant. However, the novelty of
finite superfield-dependent version of 
 BRST symmetry is that it leaves only effective action of the theory symmetric but
 not the functional measure. Remarkably, we have found that
with appropriate choices of transformation parameter this finite superfield-dependent BRST 
transformation  switches the 
void functional from one gauge-fixing condition to another gauge-fixing condition.
We have established  this connection for an  arbitrary set of gauges.
To be more  specific, the  connection of linear and non-linear gauges of 
the super-group field cosmology has also been  discussed.
  These results hold at both classical and quantum levels.
 Since the different gauge choices 
correspond to the different propagators and therefore 
our formulation will be helpful in connecting 
 different propagators of super-group field cosmology also.
Further generalizations of nilpotent BRST symmetry are the  subject of future investigations which might have some interesting implications.

\end{document}